\documentclass[12pt]{article}
\pdfoutput=1
\usepackage{epsfig,amssymb,amsmath,psfrag,multirow,epstopdf,color,graphicx,caption,subcaption}
\usepackage{breqn}
\usepackage{placeins}

\newcommand{\be}{\begin{equation}}
\newcommand{\ee}{\end{equation}}
\newcommand{\bea}{\begin{eqnarray}}
\newcommand{\eea}{\end{eqnarray}}
\newcommand{\eqn}[1]{eq.~(\ref{#1})}
\def\fig#1{fig.~{\ref{#1}}}
\def\Fig#1{Fig.~{\ref{#1}}}
\newcommand{\lr}{\leftrightarrow}
\newcommand{\Ord}{{\cal O}}

\begin{document}


\thispagestyle{empty}

\null\vskip-10pt \hfill
\begin{minipage}[t]{42mm}
SLAC--PUB--16008\\
\end{minipage}
\vspace{10mm}

\begingroup\centering
{\Large\bfseries\mathversion{bold}
Bootstrapping six-gluon scattering\\ 
in planar ${\cal N}=4$ super-Yang-Mills
theory\footnote{Talk presented by LD at 
{\it Loops \& Legs in Quantum Field Theory}, 27 April -- 2 May 2014,
Weimar, Germany}
\par}

\vspace{7mm}

\begingroup\scshape\large
Lance~J.~Dixon$^{(1)}$, James~M.~Drummond$^{(2,3,4)}$,\\
Claude Duhr$^{(5)}$, Matt von Hippel$^{(6)}$\\ and Jeffrey Pennington$^{(1)}$\\
\endgroup
\vspace{5mm}
\begingroup\small
$^{(1)}$\emph{SLAC National Accelerator Laboratory,
Stanford University, Stanford, CA 94309, USA}\\
$^{(2)}$\emph{School of Physics \& Astronomy, Univ. of Southampton,
Highfield, Southampton, SO17 1BJ, U.K.}\\
$^{(3)}$\emph{CERN, Geneva 23, Switzerland}\\
$^{(4)}$\emph{LAPTH, CNRS et Universit\'e de Savoie,
F-74941 Annecy-le-Vieux Cedex, France}\\
$^{(5)}$\emph{Institute for Particle Physics Phenomenology, 
University of Durham, Durham, DH1 3LE, U.K.}\\
$^{(6)}$\emph{Simons Center for Geometry and Physics, 
Stony Brook University, Stony Brook NY 11794, USA}\\
\endgroup

\vspace{10mm}

\textbf{Abstract}\vspace{5mm}\par
\begin{minipage}{14.7cm}
We describe the {\it hexagon function bootstrap} for solving for
six-gluon scattering amplitudes in the large $N_c$ limit of 
${\cal N}=4$ super-Yang-Mills theory.  In this method, an ansatz for the
finite part of these amplitudes is constrained at the level of
amplitudes, not integrands, using boundary information.  In the
near-collinear limit, the dual picture of the amplitudes as Wilson
loops leads to an operator product expansion which has been solved
using integrability by Basso, Sever and Vieira.  Factorization of
the amplitudes in the multi-Regge limit provides additional boundary
data.  This bootstrap has been applied successfully through four loops
for the maximally helicity violating (MHV) configuration of gluon
helicities, and through three loops for the non-MHV case.
\end{minipage}\par
\endgroup


\newpage

\section{Introduction}

It has long been a dream to construct relativistic scattering
amplitudes in four dimensions directly from their analytic
structure~\cite{ELOP}.  In most theories, there has been insufficient
information to carry out this program at the level of integrated
amplitudes, although a lot of progress has been made at the level
of loop integrands.  However, ${\cal N}=4$ super-Yang-Mills theory
in the planar limit of a large number of colors is very special.
Its perturbative amplitudes appear to have uniform transcendental weight
$2L$ for the finite parts at $L$ loops.  They also have a dual
(super)conformal invariance for any number $n$ of external
gluons~\cite{DualConformal,AMStrong,Drummond2008vq}.
This symmetry fixes the form of the four- and five-gluon
amplitudes to be equal to the BDS ansatz~\cite{BDS}.
It also requires the {\it remainder function}~\cite{RemainderFunction},
which first appears for six gluons, to depend only on
a limited number of dual conformally invariant cross ratios.

At strong coupling, amplitudes resemble soap bubbles in that the
string world-sheet has minimal area in anti-de Sitter space;
the minimal-area prescription is equivalent to computing a polygonal
Wilson loop expectation value at strong coupling~\cite{AMStrong}.
This duality between amplitudes and Wilson loops also holds at weak
coupling~\cite{WilsonLoopWeak}.  It implies that
the near-collinear limits, in which two gluon momenta are almost parallel,
can be evaluated in terms of an operator product expansion (OPE)
for the Wilson loop, in terms of the excitations of a flux 
tube~\cite{Alday2010ku,Gaiotto2010fk,Gaiotto2011dt}.
Remarkably, the OPE problem can be solved exactly in the Yang-Mills
coupling using integrability~\cite{BSVI,BSVII,BSVIII,BSVIV}.
Also, the multi-Regge limits, in which four outgoing gluons are
widely separated in rapidity, have a factorization
structure~\cite{Bartels2008ce,Bartels2008sc,Lipatov2010qg,Lipatov2010ad,%
Bartels2010tx,R63Symbol,Fadin2011we,Lipatov2012gk,Dixon2012yy}
which is ordered logarithmically and allows for the recycling
of lower-loop information to higher loops.
Finally, the super-Wilson-loop
correspondence~\cite{MasonSkinnerSuperWilson,CaronHuotSuperWilson}
leads to a set of first-order differential
equations~\cite{BullimoreSkinner,CaronHuotHe}.
These differential equations determine the full $S$ matrix in principle,
given the knowledge of higher-point, lower-loop amplitudes that
appear as source terms.  They can also be used to infer
additional, global constraints on amplitudes with a fixed number
of legs.

Perturbative amplitudes in planar ${\cal N}=4$ super-Yang-Mills theory
certainly enjoy many other fascinating properties.  However,
as we will sketch in these proceedings,
the three properties just mentioned --- the near-collinear limits,
the multi-Regge limits, and the constraints from the super-Wilson-loop ---
are extremely powerful.  Together with a specific functional
ansatz and some simpler constraints, they uniquely determine the six-gluon
MHV amplitude through at least four loops~\cite{R63Symbol,R63,R64}, and
the six-gluon non-MHV or next-to-MHV (NMHV) amplitude through at least three
loops~\cite{NMHV2L,NMHV3L}.

\section{Preliminaries}

We work with finite quantities from the very beginning:
the remainder function~\cite{RemainderFunction} for the MHV amplitude
and the ratio function~\cite{RatioFunction} for the NMHV amplitude.
In this way, we bypass all subtleties of infrared regularization.
The six-gluon remainder function $R_6$ is defined from the MHV amplitude
by factoring off the BDS ansatz~\cite{BDS},
\be
A_6^{\rm MHV}(\epsilon; s_{ij}) = 
A_6^{\rm BDS}(\epsilon; s_{ij}) \times \exp[R_6(u,v,w)] \,,
\label{R6def}
\ee
where $s_{ij}=(k_i+k_j)^2$ are momentum invariants.
This procedure not only removes all infrared divergences
(poles in $\epsilon=(4-D)/2$), it also removes a dual conformal
anomaly~\cite{Drummond2007au} in the full amplitude.  The
remainder function $R_6$ is infrared finite, and depends on just
three {\it dual conformal cross ratios}, $u$, $v$ and $w$.

The amplitude appearing in \eqn{R6def} is color-ordered.
The cyclic ordering of the six external gluons
makes it possible to define dual or sector variables $x_i$
whose differences are the momenta $k_i$:
$x_{i,i+1} \equiv x_i - x_{i+1} = k_i$.
The key transformation needed to ensure dual conformal invariance
is an inversion of the $x_i$:
\be
x_i^\mu \to \frac{x_i^\mu}{x_i^2} \,, \qquad
x_{ij}^2 \to \frac{x_{ij}^2}{x_i^2 \, x_j^2} \,.
\label{DCinversion}
\ee
This transformation leaves invariant the dual conformal cross ratios
defined by
\be
u_{ijkl} = \frac{ x_{ij}^2 \, x_{kl}^2 }{ x_{ik}^2 \, x_{jl}^2 } \,.
\label{DCinvariantsgen}
\ee
Because the external momenta are massless, the differences of adjacent
$x_i$'s have vanishing invariants, $x_{i,i+1}^2 = k_i^2 = 0$.
Consequently, there are no dual conformal cross ratios for four or five
external gluons.

\begin{figure}
\center{\includegraphics[scale=0.6,clip]{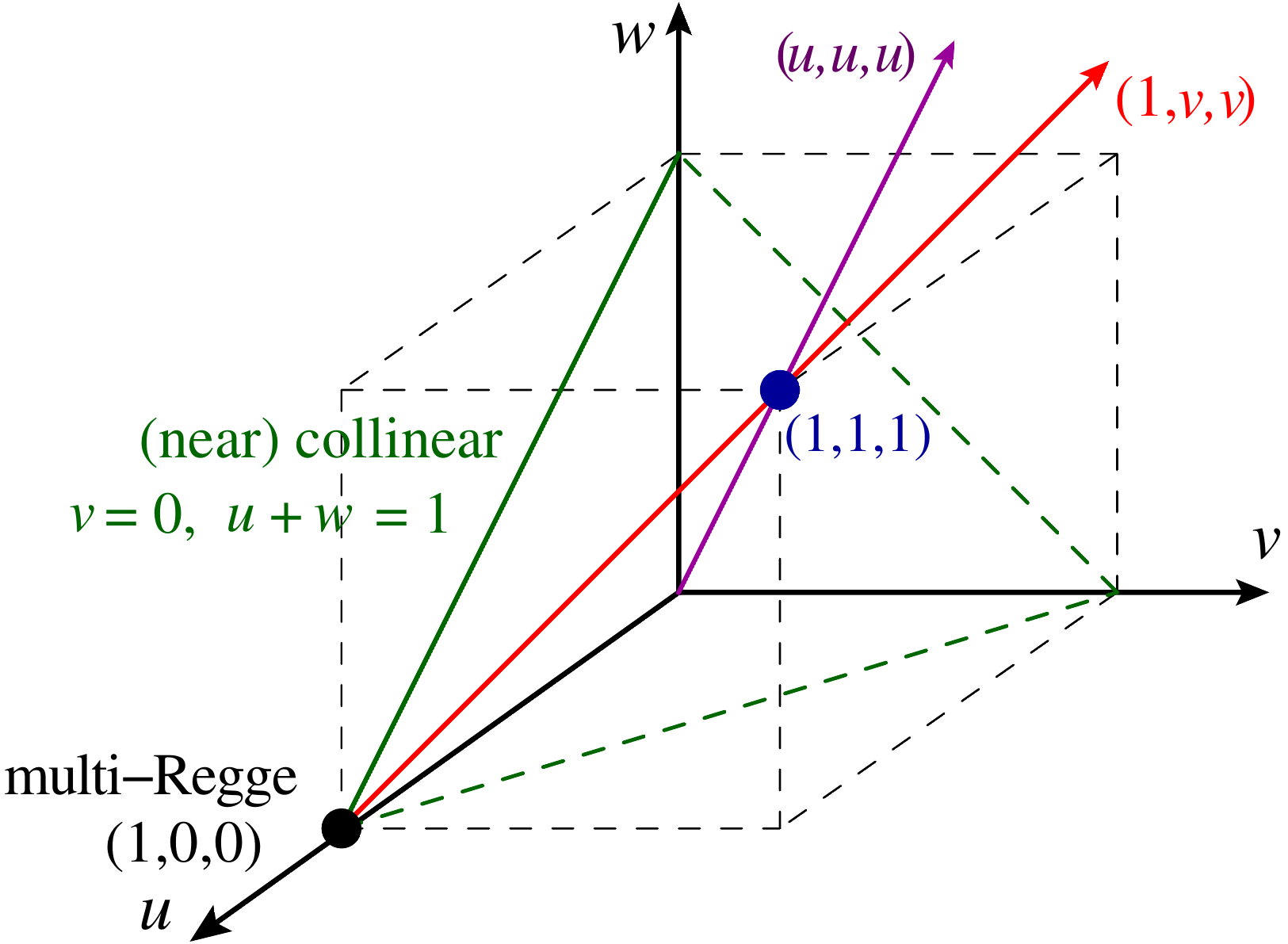}}
\caption{Space of cross ratios $(u,v,w)$ for six-gluon scattering.
Various limits are described in the text.}\label{uvwfig} 
\end{figure}

For the six-gluon case there are just three possible cross ratios,
which are related to each other by cyclic permutations:
\be
\label{eq:uvw_def}
u = \frac{x_{13}^2\,x_{46}^2}{x_{14}^2\,x_{36}^2}\,, 
\qquad v = \frac{x_{24}^2\,x_{51}^2}{x_{25}^2\,x_{41}^2}\,, \qquad
w = \frac{x_{35}^2\,x_{62}^2}{x_{36}^2\,x_{52}^2}\,.
\ee
The variables $(u,v,w)$ provide the arguments for $R_6$ in \eqn{R6def}.
\Fig{uvwfig} shows the space $(u,v,w)$, along with some distinguished
limiting regions.  The amplitude is real on a Euclidean sheet in the positive
octant with $u>0$, $v>0$ and $w>0$.

The near-collinear limit in which gluons 2 and 3 become parallel,
$x_{24}^2 = (k_2+k_3)^2 \to 0$, sends $v\to0$ and also constrains $u+w$ to be 1.  
It is shown as the solid green line in the figure.
There are two other limits, related by cyclic symmetry, shown
as dashed green lines. The near-collinear limit can be taken on the
Euclidean sheet.
The multi-Regge limit for $2\rightarrow4$ scattering
is obtained by first moving onto a Minkowski sheet
by taking $u\to e^{-2\pi i} u$.  Then one approaches the black point in the
figure, $(u,v,w) \to (1,0,0)$, by taking
$u\to1$ and $v,w\to0$ with the ratios $v/(1-u)$ and $w/(1-u)$ held fixed.  
These two limits provide the main physical constraints we impose on the
amplitude. The analytic expressions for the amplitudes simplify somewhat
on the red line $(1,v,v)$ and on the purple line $(u,u,u)$.
They reduce further, to linear combinations of multiple zeta values (MZVs),
at the intersection of these lines at the point $(u,v,w)=(1,1,1)$.

\section{Hexagon Functions}

We can expand the remainder function perturbatively,
\be
R_6(u,v,w) = \sum_{L=2}^\infty a^L \, R_6^{(L)}(u,v,w),
\label{R6L}
\ee
where $a=g_{\rm YM}^2N_c/(8\pi^2)$ is the 't Hooft coupling
constant, $g_{\rm YM}$ is the Yang-Mills coupling constant and
$N_c$ is the number of colors.  The remainder function
begins at two loops~\cite{RemainderFunction}.  Taking
inspiration from the simplified analytic form of the two-loop
result~\cite{GSVV}, we assume~\cite{R63Symbol,R63}
that the $L$-loop remainder function
$R_6^{(L)}$ is a linear combination of weight-$2L$ 
{\it hexagon functions} (to be defined shortly).

The NMHV amplitude really consists of many different components,
related to each other by ${\cal N}=4$ supersymmetry.
It is best described using an
on-shell superspace~\cite{Nair} and defining the ratio of the NMHV
superamplitude to the MHV superamplitude~\cite{Drummond2008vq},
\be
{\cal P}^{\rm NMHV} \equiv \frac{{\cal A}^{\rm NMHV}}{{\cal A}^{\rm MHV}} \,,
\label{PNMHV}
\ee
We assume~\cite{NMHV2L} that this super-ratio can be expressed as
\bea
{\cal P}^{\rm NMHV} &=& \frac{1}{2}\Bigl[
 [(1) + (4)] V(u,v,w) + [(2) + (5)] V(v,w,u) + [(3) + (6)] V(w,u,v)  \\
&&\hskip0.0cm 
+ [(1) - (4)] \tilde{V}(u,v,w) - [(2)-(5)] \tilde{V}(v,w,u)
  + [(3) - (6)] \tilde{V}(w,u,v) \Bigr] \,,
\label{PVform}
\eea
where $(1) \equiv [23456]$ is a shorthand for a certain
five-bracket dual superconformal $R$-invariant, and where
the $L$-loop coefficients in the perturbative expansions of
$V$ and $\tilde{V}$,
\bea
V(u,v,w) &=& \sum_{L=0}^\infty a^L \, V^{(L)}(u,v,w),
\label{VL}\\
\tilde{V}(u,v,w) &=& \sum_{L=2}^\infty a^L \, \tilde{V}^{(L)}(u,v,w),
\label{VtL}
\eea
are also weight-$2L$ hexagon functions.

Hexagon functions form a certain class of iterated
integrals~\cite{Chen} or multiple polylogarithms~\cite{FBThesis,Gonch}.
If we differentiate a weight-$n$ function $f$ in this class, the result
can be written as
\be
df = \sum_{s_k\in {\cal S}} f^{s_k} \, d \ln s_k \,,
\label{df}
\ee
where ${\cal S}$ is a finite set of rational expressions, known as
the letters of the symbol, and $f^{s_k}$ are weight-$(n-1)$ functions in
the same class.  These functions describe the $\{n-1,1\}$ component
of a coproduct $\Delta$ associated with a Hopf algebra for
iterated integrals~\cite{Gonch3,Gonch2,Brown2011ik}.
Similarly, we can differentiate each $f^{s_k}$,
\be
df^{s_k} = \sum_{s_j\in {\cal S}} f^{s_j \, s_k} \, d \ln s_j \,,
\label{dfk}
\ee
thereby defining the weight-$(n-2)$ functions $f^{s_j \, s_k}$, which describe
the $\{n-2,1,1\}$ components of $\Delta$.  The maximal iteration
of this procedure defines the symbol of $f$, an $n$-fold tensor product
of elements of ${\cal S}$ (each standing for a $d\ln$).

Hexagon functions are functions whose symbols have letters drawn from
the nine-letter set,
\be
{\cal S} = \{u,v,w,1-u,1-v,1-w,y_u,y_v,y_w\} \,.
\label{Sletters}
\ee
The nine letters can be understood to arise from momentum
twistors~\cite{Hodges} $Z_i^A$, $i=1,2,\ldots,6$, $A=1,2,3,4$,
because these objects transform simply under dual conformal
transformations, as do the four-brackets
$\langle ijkl\rangle \equiv \varepsilon_{ABCD} Z_i^A Z_j^B Z_k^C Z_l^D$.
However, the four-brackets are not invariant under projective
transformations (rescalings).  There are 15 projectively invariant
ratios of four-brackets, and they can be factored into the nine
basic ones, given in \eqn{Sletters}.
The three variables $y_u$, $y_v$, $y_w$ are not independent of $u,v,w$
but satisfy
\be
y_u = \frac{u-z_+}{u-z_-}\,, \qquad y_v = \frac{v-z_+}{v-z_-}\,, 
\qquad y_w = \frac{w - z_+}{w - z_-}\,,
\label{yfromu}
\ee
where
\be
z_\pm = \frac{1}{2}\Bigl[-1+u+v+w \pm \sqrt{\Delta}\Bigr]\,, 
\qquad \Delta = (1-u-v-w)^2 - 4 uvw\,.
\label{zDeltadef}
\ee
Although $y_u,y_v,y_w$ contain square roots when expressed in terms
of $u,v,w$, the converse is not true; $u$ is given by the rational
expression,
\be
u = \frac{y_u (1 - y_v) (1 - y_w)}{(1 - y_u y_v) (1 - y_u y_w)}\,,
\label{u_from_y}
\ee
and $v$ and $w$ are given by cyclic permutations of this relation.

Hexagon functions have one other defining property:
Their branch cuts should start only at the origin in the
momentum invariants $(k_i+k_{i+1})^2$ and $(k_i+k_{i+1}+k_{i+2})^2$,
which means they should start only at 0 or $\infty$ in the
cross ratios $u,v,w$.   In terms of the symbol, the first entry must
be drawn from $\{u,v,w\}$~\cite{Gaiotto2011dt}.

In certain regions, hexagon functions collapse to simpler functions.
For example:
\begin{enumerate}
\item On the lines $(u,u,1)$ and $(u,1,1)$, the set ${\cal S}$ collapses to
$\{u,1-u\}$, corresponding to the {\it harmonic} polylogarithms (HPLs)
defined by Remiddi and Vermaseren~\cite{HPL}, with weight vectors
$0$ and $1$ only.
\item On the line $(u,u,u)$, ${\cal S}$ collapses to
$\{y_u,1+y_u,1+y_u+y_u^2\}$, corresponding to {\it cyclotomic}
polylogarithms~\cite{Cyclotomic}, because $1+y_u$ and $1+y_u+y_u^2$
vanish when $y_u$ is a sixth root of unity.
\item In the multi-Regge limit, with $v/(1-u) = 1/[(1-z)(1-\bar{z})]$,
$w/(1-u) = z\bar{z}/[(1-z)(1-\bar{z})]$, the relevant
functions~\cite{Dixon2012yy}
have symbol letters $\{z,1-z,\bar{z},1-\bar{z}\}$, plus a first-entry
or single-valuedness constraint.  These {\it single-valued harmonic}
polylogarithms (SVHPLs) have been studied by Brown~\cite{BrownSVHPLs}.
\end{enumerate}

Two complementary methods have been used to construct hexagon
functions~\cite{R63,R64}.  One approach is to represent each hexagon
function in terms of multiple polylogarithms in the $y_i$ variables,
for a particular region in the space of cross ratios.
The multiple polylogarithms can be evaluated numerically using
{\sc GiNaC}~\cite{Bauer2000cp,Vollinga2004sn}.
They can also be expanded analytically in various limits as
needed to impose physical constraints.   Alternatively,
one can define the hexagon functions iteratively via the $\{n-1,1\}$
components of their coproducts, which amount to a set of
coupled first-order differential equations.  These equations
can be integrated numerically, or solved analytically in special
limits, such as the ones enumerated above.

The complete set of hexagon functions through weight five has been
described using both methods~\cite{R63}.
This set suffices to characterize the weight-six three-loop
functions $R_6^{(3)}$, $V^{(3)}$ and $\tilde{V}^{(3)}$ via their
weight-five $\{5,1\}$ coproduct components, {\it i.e.}~their first
derivatives, up to constants of integration which can also be fixed.
To go to four loops, one can start by enumerating the possible
$\{5,1,1,1\}$ coproduct components, and then work upward, through
the $\{6,1,1\}$ and $\{7,1\}$ components, in order to characterize the
full space of weight-eight functions.  Some mathematical consistency
conditions have to be imposed throughout the construction, namely the
equality of mixed partial derivatives, and the absence of branch cuts
in undesired locations.

\section{Applying constraints}

Once one understands the space of functions, the basic strategy of the
hexagon function bootstrap at $L$ loops is very simple:
\begin{enumerate}
\item Enumerate all the hexagon functions at weight $2L$.  This list
contains products of lower-weight functions, such as the products of
weight-$k$ MZVs (constants) with weight-$(2L-k)$ hexagon functions.
\item Write the most general linear combination of such functions
with unknown rational-number coefficients.
\item Impose certain symmetry properties and other simple constraints,
followed by the super-Wilson-loop, OPE and multi-Regge constraints,
until all coefficients have been uniquely determined.
\end{enumerate}
Sometimes we perform the last step in two stages.  In the first stage,
we fix the symbol of the desired function.  In the second stage, we determine
the full function, including some Riemann $\zeta$-valued ambiguities
present at symbol level.

The simple constraints on $R_6$ are:
\begin{enumerate}
\item Total symmetry under exchange of $u,v,w$.
\item Even under ``parity'' ($y_i \lr 1/y_i$); every term in its symbol
must contain an even number of $y_i$.
\item Vanishing in the collinear limit, $R_6\to0$ as $v\to0$, $u+w\to1$.
\item Final entry restricted~\cite{CaronHuotHe}
 to six combinations: $\{u/(1-u),v/(1-v),w/(1-w),y_u,y_v,y_w\}$.
\end{enumerate}
The simple constraints on $V(u,v,w)$ ($\tilde{V}(u,v,w)$) are:
\begin{enumerate}
\item Symmetric (antisymmetric) under exchange of $u$ and $w$.
\item Even (odd) under ``parity''.
\item Cancellation of spurious poles arising from the prefactors $(i)$.
\item Collinear vanishing.
\item Final entry restricted~\cite{CaronHuotHe,SimonPrivate}
 to seven combinations, where the extra combination is $uw/v$.
\end{enumerate}
The spurious-pole and collinear constraints on the ratio function involve
a couple of different permutations of $V$ and $\tilde{V}$~\cite{NMHV2L}.

The OPE constraints are imposed in the limit $\tau\to\infty$, where
$v \approx T^2 \to 0$, $T \equiv e^{-\tau}$.  The other relevant variables
in this limit are $\sigma\approx \frac{1}{2} \ln(u/w)$, which characterizes
the longitudinal splitting fraction of the two collinear gluons; and $\phi$,
which is an azimuthal angle.
The original OPE constraints~\cite{Alday2010ku,Gaiotto2010fk,Gaiotto2011dt}
only fixed the ``leading discontinuity'' terms, which for the remainder
function have the form $(\ln T)^{L-1}$.
The new integrability-based results give all powers of $\ln T$
for the leading power-law (leading twist) behavior coming from
a single flux-tube excitation~\cite{BSVI,BSVII}.  These terms
have the form $T e^{\pm i\phi} (\ln T)^k f_k(\sigma)$, $k=0,1,2,\ldots,L-1$,
for some functions $f_k(\sigma)$ involving HPLs.
More recently, the two-excitation contributions have also become
available~\cite{BSVIII}; they behave like
$T^2 \{ e^{\pm 2i\phi},1\} (\ln T)^k f_k(\sigma)$, $k=0,1,2,\ldots,L-1$.
The $\phi$ dependence is controlled by the angular momentum of
the excitations.  At leading twist, only gluons (spin $\pm1$) contribute.
At the next order, pairs of fermions and scalars can also contribute,
to the 1 term but not to the $e^{\pm 2i\phi}$ terms, which are purely gluonic.
The gluonic terms can be controlled to arbitrary order in $T$~\cite{BSVIV}.

The imposition of the multi-Regge constraints is described
elsewhere~\cite{R63Symbol,Dixon2012yy,R63,R64,NMHV3L}.

In the case of the remainder function, we fixed the symbol first.
Table~\ref{multi_loop_symbol} summarizes how many unknown parameters
are left after each constraint is imposed.
There are also parameters multiplying Riemann
$\zeta$ values, to which the symbol is insensitive.  At four loops,
for example, there are 68 such parameters~\cite{R64}.  They can be fixed
by imposing the same constraints at function level.
For the $T^2$ OPE constraints, only the $T^2 \cdot e^{\pm2i\phi}$ terms had
to be imposed to fix everything.  The $T^2 \cdot 1$ terms provided
a pure cross check.

\renewcommand{\arraystretch}{1.25}
\begin{table}[!ht]
\begin{center}
\begin{tabular}{|l|c|c|c|}
\hline\hline
\multicolumn{1}{|c|}{Constraint} &\multicolumn{1}{c|}{$L=2$}
&\multicolumn{1}{c|}{$L=3$} &\multicolumn{1}{c|}{$L=4$} \\
\hline\hline
1. Integrability & 75 & 643 & 5897  \\
\hline
2. Total $S_3$ symmetry & 20 & 151 & 1224 \\
\hline
3. Parity invariance & 18 & 120 & 874 \\
\hline
4. Collinear vanishing ($T^0$) & 4 & 59 & 622 \\
\hline
5. OPE leading discontinuity & 0 & 26 & 482 \\
\hline
6. Final entry & 0 & 2 & 113  \\
\hline
7. Multi-Regge limit & 0 & 2 & 80 \\
\hline
8. Near-collinear OPE ($T^1$) & 0 & 0 & 4\\
\hline
9. Near-collinear OPE ($T^2$) & 0 & 0 & 0\\
\hline\hline
\end{tabular}
\caption{\label{multi_loop_symbol} Remaining parameters
in the symbol of $R_6^{(L)}$ at loop order $L=2,3,4$, after
applying the various constraints successively~\cite{R64}.
Once a ``0'' appears, the symbol is uniquely determined.  Further 0's
represent cross checks from additional constraints.}
\end{center}
\end{table}

For the ratio function, we imposed the constraints at the function level
from the beginning.  The number of parameters remaining in this
case is shown in table~\ref{NMHV_constr}.  The cyclic vanishing
constraint on $\tilde{V}$ is required because of an identity obeyed
by the $(i)$ invariants, $[(1)+(3)+(5)]-[(2)+(4)+(6)] = 0$.
For details on how the remaining constraints were imposed,
see refs.~\cite{NMHV2L,NMHV3L}.  Either the $T^2$ OPE or the multi-Regge
constraints were sufficient to fix the last two parameters, leaving
the other set of constraints as a pure cross check.

\renewcommand{\arraystretch}{1.25}
\begin{table}[!t]
\centering
\begin{tabular}[t]{|l|c|c|c|}
\hline\hline
\multicolumn{1}{|c|}{Constraint} &\multicolumn{1}{c|}{$L=1$}
&\multicolumn{1}{c|}{$L=2$} &\multicolumn{1}{c|}{$L=3$} \\
\hline\hline
1. (Anti)symmetry in $u$ and $w$ & 7 & 52 & 412\\\hline
2. Cyclic vanishing of $\tilde{V}$ & 7 & 52 & 402\\\hline
3. Final-entry condition & 4 & 25 & 182\\\hline
4. Spurious-pole vanishing & 3 & 15 & 142\\\hline
5. Collinear vanishing & 1 & 8 & 92\\\hline
6. $\Ord(T^1)$ OPE & 0 & 0 & 2\\\hline
7. $\Ord(T^2)$ OPE {\it or} multi-Regge kinematics & 0 & 0 & 0\\
\hline\hline
\end{tabular}
\caption{Remaining parameters in the function-level
ans\"{a}tze for $V^{(L)}$ and $\tilde{V}^{(L)}$ after each constraint is applied,
at each loop order~\cite{NMHV3L}.}
\label{NMHV_constr}
\end{table}

\section{Results}

Having determined the functions $R_6$, $V$ and $\tilde{V}$,
we can plot them, and examine some of their analytic properties.
At the point $(u,v,w)=(1,1,1)$, hexagon functions all collapse to
MZVs.  For the remainder function we find,
\bea
R_6^{(2)}(1,1,1) &=& - (\zeta_2)^2 = - \frac{5}{2} \zeta_4 \,, \\
R_6^{(3)}(1,1,1) &=& \frac{413}{24} \, \zeta_6 + (\zeta_3)^2 \,, \\
R_6^{(4)}(1,1,1) &=& -\frac{3}{2}\zeta_2(\zeta_3)^2 - \frac{5}{2}\zeta_3 \zeta_5
- \frac{471}{4} \zeta_8 + \frac{3}{2}\zeta_{5,3} \,.
\eea
The first MZV that is irreducible (cannot be written in terms of $\zeta_k$'s)
occurs at weight eight; this quantity, $\zeta_{5,3}$,
does appear in $R_6^{(4)}(1,1,1)$.

\begin{figure}
\begin{center}
\includegraphics[width=6in]{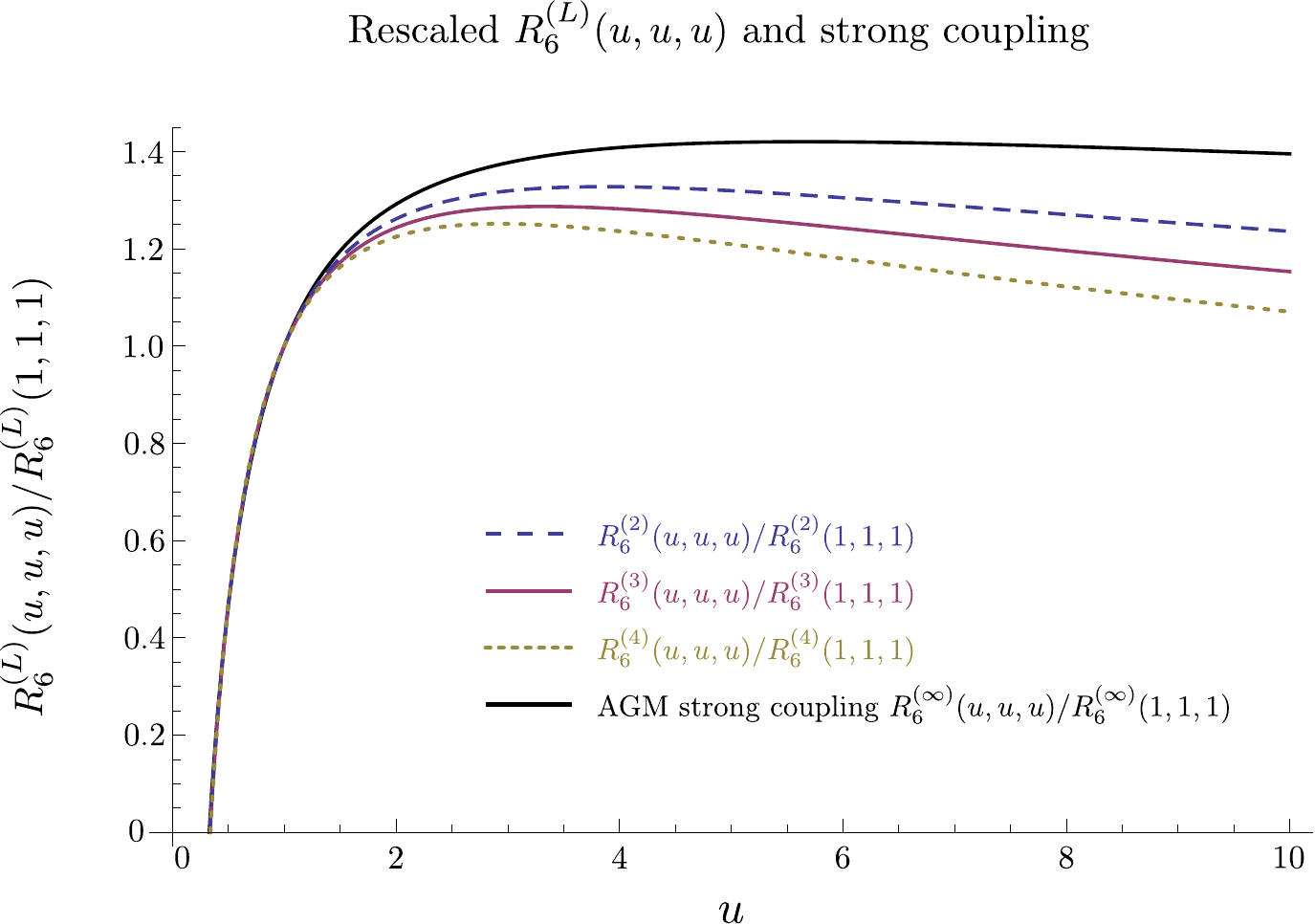}
\end{center}
\caption{The remainder function on the line $(u,u,u)$, plotted at two,
three, and four loops and at strong coupling, after normalizing the
coefficient functions by their values at the point
$(1,1,1)$~\cite{R64}.}
\label{uuu_agm_R64}
\end{figure}

An interesting line on which to plot the remainder function
is when all three cross ratios are equal, {\it i.e.}~the line $(u,u,u)$.
\Fig{uuu_agm_R64} shows the two-, three-, four-loop
and strong-coupling remainder functions on this line.
The strong-coupling result is computed from the minimal-area
prescription~\cite{AMStrong,Alday2009dv}.\footnote{%
Recently another strong-coupling contribution was
identified~\cite{BSVStrong}.  It is a constant in $(u,v,w)$,
not included here.}
The plots are not partial sums --- that would require us to choose
a value for the 't Hooft coupling --- but rather the individual coefficient
functions.  We have rescaled them all by their values at $u=1$.
Remarkably, for $u<1$, all four curves have very similar shapes,
even though they are composed of quite different analytic functions
--- cyclotomic polylogarithms~\cite{Cyclotomic} of different weights
at weak coupling, and an arccosine at strong coupling~\cite{Alday2009dv}.
This remarkable similarity in shape was noticed already at two
loops~\cite{Hatsuda2012pb}.\footnote{See
refs.~\cite{Brandhuber2009da,DelDuca2010zp,Hatsuda} for similar
observations for other kinematical configurations.}
It clearly persists through four loops for $u<1$.
More generally, ratios of successive loop orders for the remainder
function tend to vary quite slowly with $u,v,w$, at least within the
unit cube and when $u,v,w$ are not too small.

\begin{figure}
\begin{center}
\includegraphics[width=4.0in]{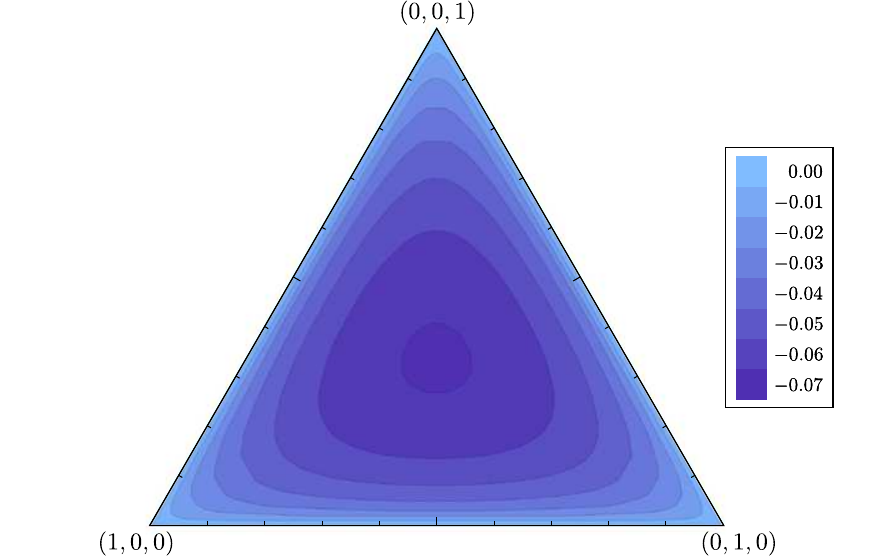}\\
\includegraphics[width=3.6in]{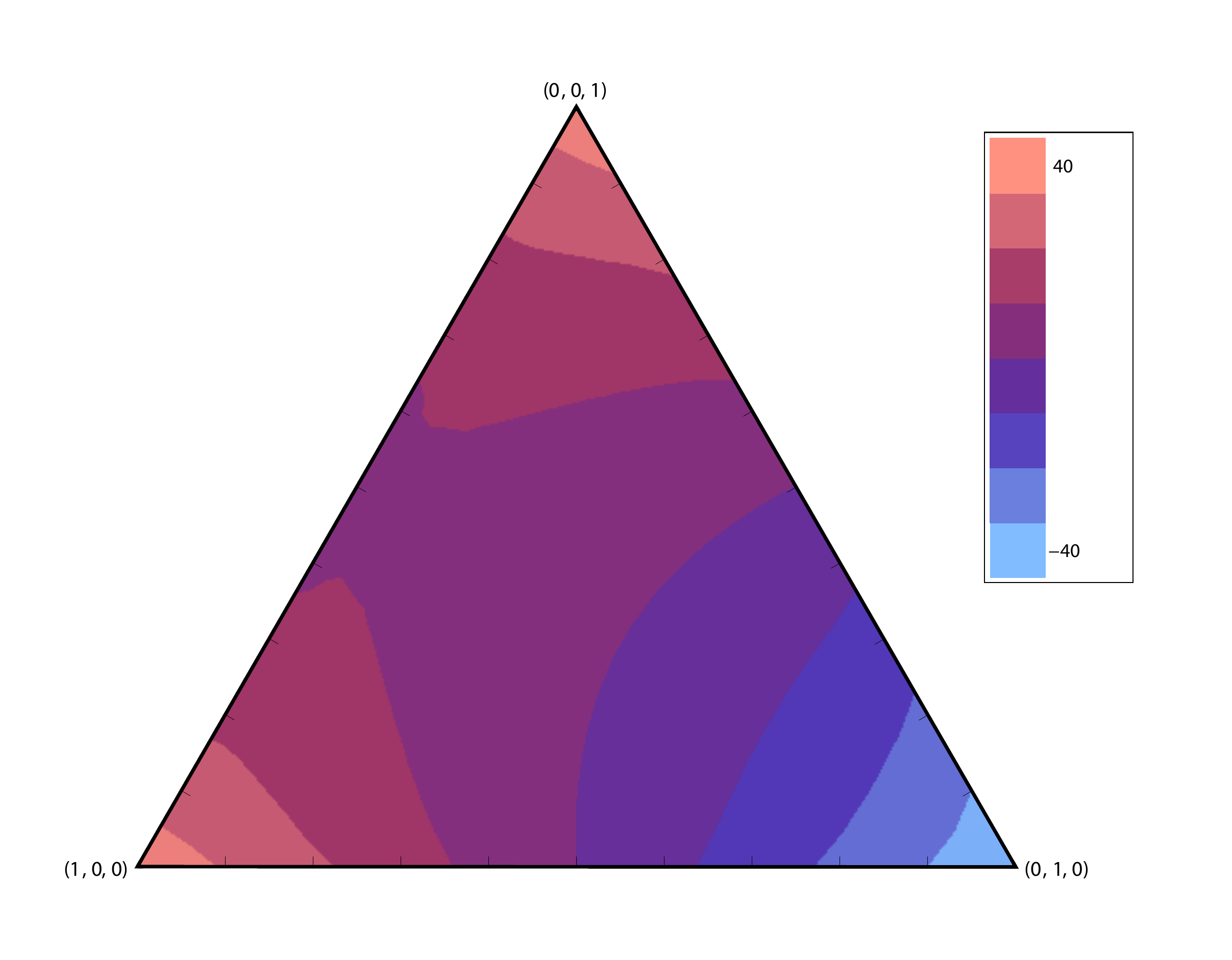}\hskip-1.0cm
\includegraphics[width=3.6in]{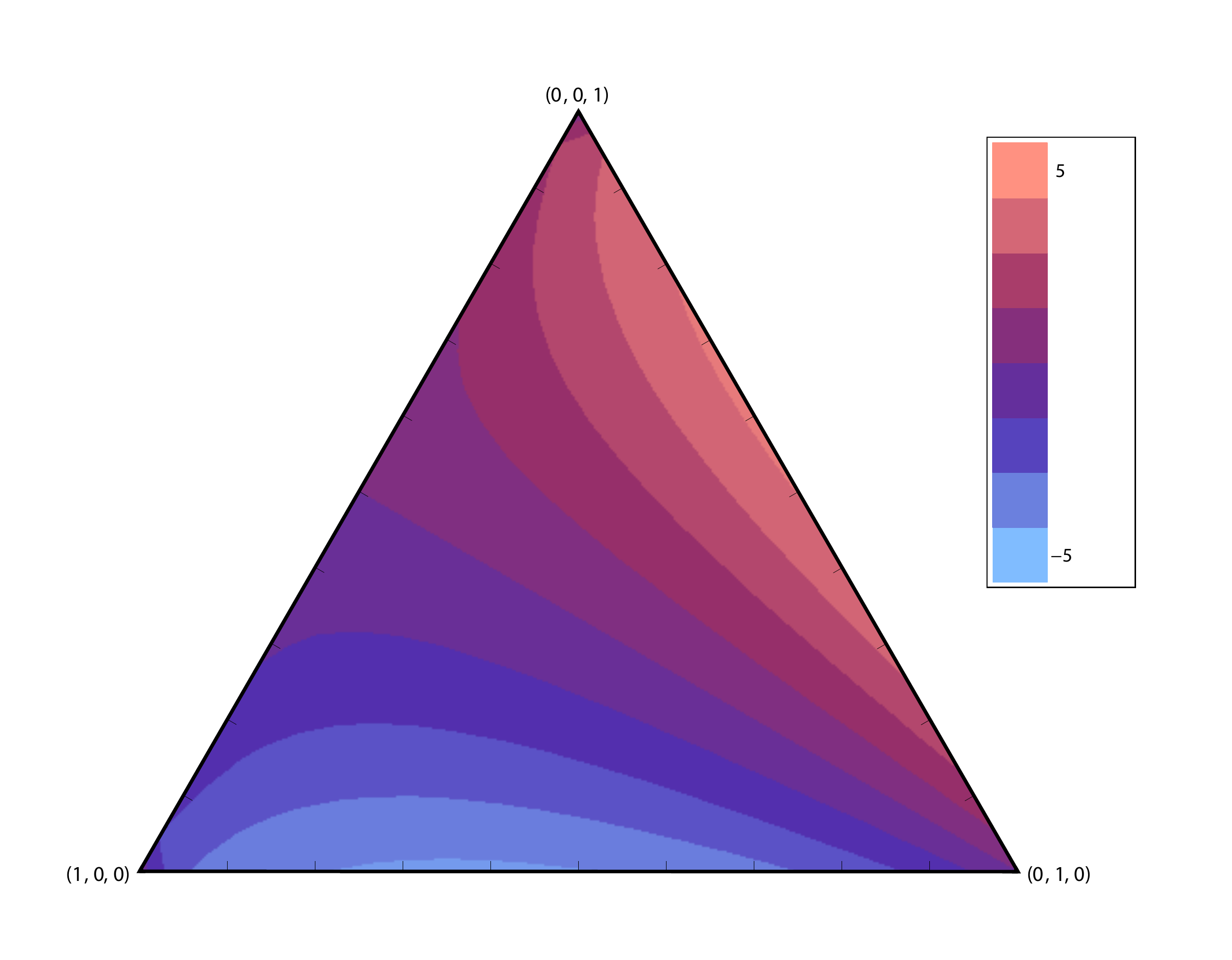}
\end{center}
\caption{The three-loop functions $R_6^{(3)}$, $V^{(3)}$ and $\tilde{V}^{(3)}/i$,
plotted on the plane $u+v+w=1$~\cite{R63,NMHV3L}.}
\label{triangleplots}
\end{figure}

Finally, in \fig{triangleplots}
we plot the three-loop remainder function
$R_6^{(3)}(u,v,w)$ and ratio function components $V^{(3)}(u,v,w)$
and $\tilde{V}^{(3)}(u,v,w)$ on the triangle which is
the part of the plane $u+v+w=1$ that lies in the positive octant.
The remainder function is constrained to vanish on all three edges of
the triangle (the collinear limits); consequently it never gets very large
in the interior.  For the ratio function, the collinear-vanishing constraint
involves a linear combination of two permutations of $V$, and so $V^{(3)}$
does not have to vanish on the edges.  ($\tilde{V}^{(3)}$, like
all parity-odd hexagon functions, actually does
vanish on the edges, but the vanishing happens so slowly that it is
not visible in the plots.)

\vfill\eject

\section{Conclusions}

Scattering amplitudes in planar ${\cal N}=4$ super-Yang-Mills theory
are so strongly constrained that it is feasible to compute them using
a bootstrap that operates at the level of integrated amplitudes,
without any direct knowledge of the integrands at high loop orders.
We have demonstrated the practicality of this bootstrap for the six-gluon
scattering amplitude, for both the MHV and NMHV helicity configurations.
Of course, some notion of the space in which the solution lies is necessary.
In the six-gluon case it is provided by hexagon functions, motivated by
the simple form found earlier for the two-loop remainder
function.  The existence of detailed
boundary value ``data'' from the near-collinear and multi-Regge limits
was also essential.  The numerical and analytic results obtained so far
exhibit interesting patterns, which may help in constructing all-orders
solutions.  It would be very interesting to try to extend these methods
to other amplitudes in planar ${\cal N}=4$ super-Yang-Mills theory,
and to more general theories as well.


\vskip0.4cm

{\bf Acknowledgments}
This research was supported by the US Department of Energy under
contracts DE--AC02--76SF00515 and DE--FG02--92ER40697, by the
Research Executive Agency (REA) of the European Union under the Grant
Agreement PITN-GA-2010-264564 (LHCPhenoNet) and by the
EU Initial Training Network in High-Energy Physics and Mathematics: GATIS. 
LD is grateful to the organizers of {\it Loops and Legs} for the
opportunity to present this work at such a stimulating meeting,
and to the Simons Center for Geometry and Physics for hospitality
when some of this research was performed.


\end{document}